\documentclass[journal=jpclcd,layout=twocolumn]{achemso}
\usepackage[utf8]{inputenc}
\usepackage{xcolor}
\usepackage{geometry}
\usepackage{graphicx}
\usepackage[normalem]{ulem}
\newcounter{Requ}

\newcommand{\be}{\begin{equation}}
\newcommand{\ee}{\end{equation}}
\title{High capacity NbS$_2$-based anodes for Li-ion batteries}
\author{Alexandra Carvalho}
\affiliation{Institute for Functional Intelligent Materials, National University of Singapore, 117544, Singapore}
\email{carvalho@nus.edu.sg}
\author{Vivek Nair}
\affiliation{ CRADLE Singapore, Hyundai Motor Innovation Center in Singapore}
\author{Sergio G. Echeverrigaray}
\affiliation{Centre for Advanced 2D Materials, National University of Singapore, 117546 Singapore}
\email{sergio@nus.edu.sg}
\author{Antonio H. Castro Neto}
\altaffiliation{Institute for Functional Intelligent Materials, National University of Singapore, 117544, Singapore}
\alsoaffiliation[Second University]
{Centre for Advanced 2D Materials, National University of Singapore, 117546 Singapore}
\affiliation{Department of Materials Science Engineering, National University of Singapore, 117575 Singapore}

\keywords{Batteries, Niobium sulfide, NbS$_2$, cathode, anode, capacity, oxygen, vacancies, defects, intercalation, conversion, alloy}

\begin{tocentry}
\includegraphics[width=5cm]{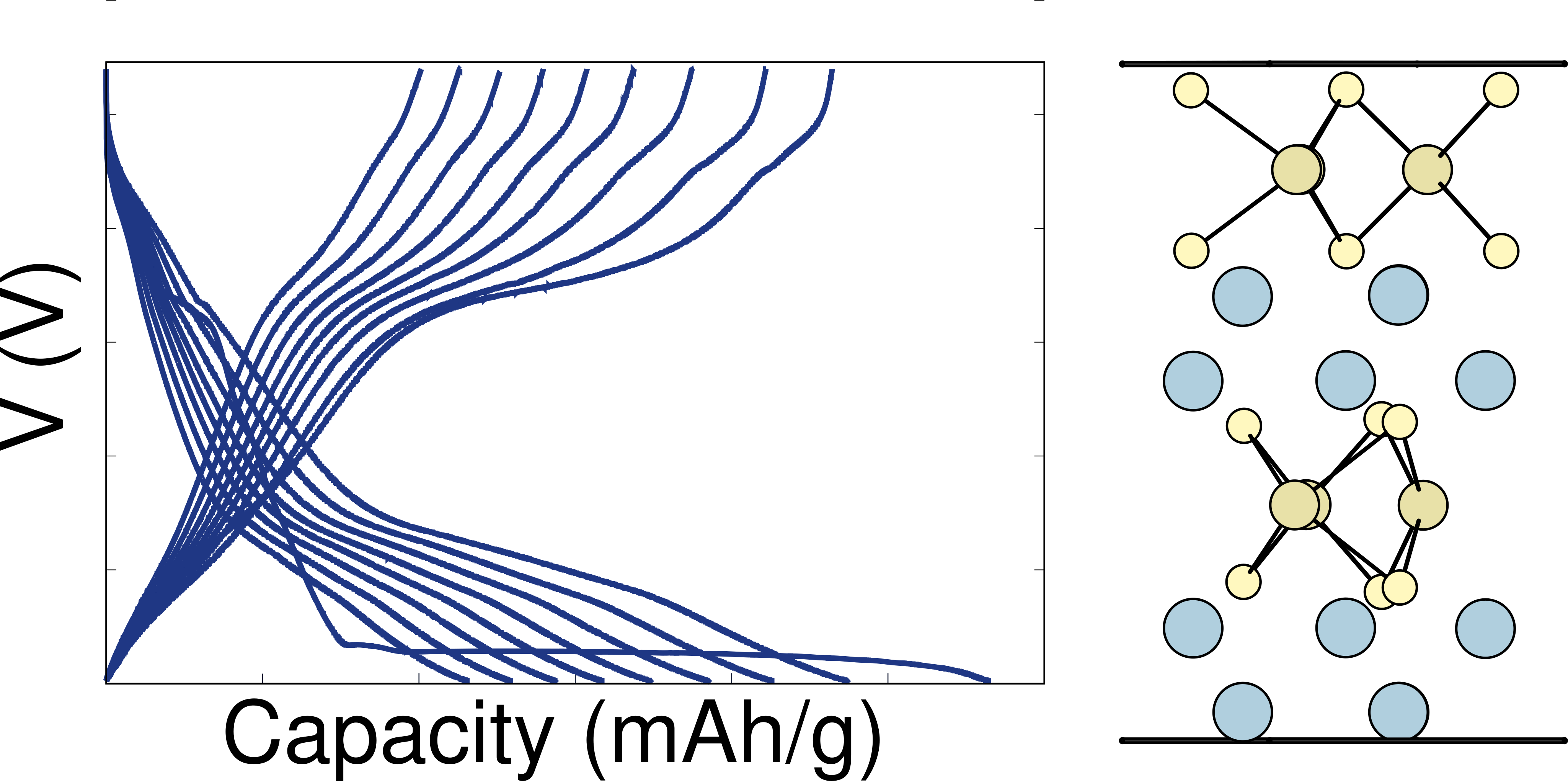}
\end{tocentry}
\begin{document}

\maketitle
\begin{abstract}
We have investigated the lithium capacity of the 2$H$ phase of niobium sulphide (NbS$_2$) using  density functional theory calculations and experiments. Theoretically, this material is found to allow the intercalation of a double layer of Li in between each NbS$_2$ layer when in equilibrium with metal Li. The resulting specific capacity (340.8~mAh/g for the pristine material, 681.6 mAh/g for oxidized material) can reach more than double the specific capacity of graphite anodes. The presence of various defects leads to an even higher capacity with a partially reversible conversion of the material, indicating that the performance of the anodes is robust with respect to the presence of defects. Experiments in battery prototypes with NbS$_2$-based anodes find a first specific capacity of about 1,130 mAh/g, exceeding the theoretical predictions. 
\end{abstract}

Mobile technologies have become an increasingly pervasive part of daily life, and have led to an insatiable consumer demand for smaller, higher-capacity, fast-charging batteries. Concurrently, vehicle electrification is driving a demand for batteries that allow for a longer vehicle range, but at the same time have a longer lifespan (cyclability), and are safer and environmentally friendly.

Currently, most of the existing Li-ion batteries use an intercalation-type anode material like graphite. Graphite provides a specific capacity of 372~mAh/g\cite{asenbauer2020success} but presents various problems like dendrite formation and plating which cause irreversible capacity loss and safety hazards\cite{orsini1998situ}. It is expected that an improvement of the energy density of automotive cells above 260 Wh/kg ceiling can be achieved by pairing up layered nickel-rich ternary cathode materials with either Si-containing anodes\cite{schmuch2018performance} or Li-metal anodes\cite{chen2019critical}. However, these anodes also present stability issues.
Li metal has the highest theoretical Li packing density, but when it is used as an anode, it undergoes platting/stripping\cite{liu2019plating} while the cathode undergoes intercalation/de-intercalation (in the case of insertion compounds) or conversion reactions (in the case of the sulphur cathode). This results in capacity loss, dendrite formation, and stability issues.
Silicon anodes have a theoretical Li capacity of about 4200 mAh/g, being able to store 4.4 Li ions per original Si atom (Li$_{22}$Si$_5$) when fully lithiated\cite{chan2012first,cubuk2014theory}, not by intercalation, but through a ``conversion” mechanism, where silicon and lithium ions form an electrochemical alloy, breaking/restoring bonds during charge/discharge cycling. Since the bonds established in the alloy are much stronger than the typical electrostatic and van der Waals attraction between the Li ions and intercalation hosts, there is irreversible damage during the charge/discharge cycle, and therefore attaining Si anode cyclability is much more problematic\cite{jin2017challenges,nitta2015li}.
Another major challenge with silicon anodes is that they typically expand by up to 300\% of their original volume during lithiation. This swelling can lead to mechanical stresses and fractures in the anode material, which can cause capacity loss and reduced cycle life of the battery, and has so far prevented widespread adoption of pure silicon as an anode material. Various strategies have been investigated to mitigate this issue. Using nanostructured silicon materials, it is possible to cushion the expansion and prevent stress fracture and pulverization\cite{phadatare2019silicon,li2014mesoporous,liu2012size,han2021achieving}, but damage still occurs at the silicon surface, affecting the solid electrolyte interphase (SEI) and leading to capacity fade\cite{jin2017challenges}. To use silicon to boost capacity while avoiding these challenges, manufacturers add less than 10\% silicon or silicon oxide  to graphite anodes\cite{blomgren2016development,schmuch2018performance}.
It is therefore desirable to find anode materials that have high capacity but without the accompanying volume expansion that is problematic in silicon anodes, either to use as main anode material or as an additive.

Transition metal dichalcogenides can host intercalated Li, similar to graphite.\cite{chen2020transition}
However, conversion-type reactions where the dichalcogenide layer reacts with Li, similar to the lithiation of silicon,  have also been predicted theoretically.\cite{zhao2019electrochemical}

The lithium intercalation capacity can be found by density functional theory calculations or other theoretical methods, as it has been shown in the case of graphite (of which the intercalated phase is LiC$_6$ )\cite{persson2010thermodynamic,lenchuk2019comparative,raju2015reactive,li2019intercalation} 
Similarly, the capacity of conversion-type anodes has also been studied using atomistic simulations.\cite{chan2012first,cubuk2014theory}
Some studies of Li insertion in compound layered materials have intercalation stages similar to graphite\cite{liu2020density,xu2014adsorption,carvalho2014donor}.
There are several previous studies of Li adsorption on a NbS$_2$ monolayer\cite{de2021toward,zhao2016phosphorene,zhu2017functionalized}. 
A previous study of bulk 2$H$-NbS$_2$ (which we will just refer to as NbS$_2$) found that its structure is largely unchanged by lithium insertion.\cite{zhao2019electrochemical}
However, there is a need to understand the stages of lithiation of the bulk NbS$_2$ and the influence of the presence of defects on the onset of the conversion-type reactions, and the respective maximum capacity.

In this article, we show that NbS$_2$ is a high-capacity anode material, with two Li insertion phases: intercalation, up to 340.8 mAh/g, followed by a partially reversible conversion (alloying) phase dependent on the presence of defects, up to 1459.8 mAh/g. We propose that a double Li layer intercalation is responsible for a higher capacity of NbS$_2$ in pristine material. We also characterise experimentally the Galvanic charge/discharge cycling of NbS$_2$-based anodes in NbS$_2$/Li half cells, demonstrating the robustness of the high anode capacity. A comparison between theory and experiment enables us to identify the main Li insertion stages observed.

The voltage across a battery or half cell, at a point in time when the fraction of Li in the cathode is $x$, is given by:
\be V(x) = -\frac{\mu_{\rm Li}^{\rm cathode}(x)-\mu_{\rm Li}^{\rm anode}(x)}{F},\ee
where $\mu_{\rm Li}^{\rm cathode}(x)$ and $\mu_{\rm Li}^{\rm anode}(x)$ are the chemical potentials of the cathode and anode, and $F$ is the Faraday constant. An anode material is typically tested in a half-cell with a body-centered cubic (bcc) Li metal electrode, and in that case $V(x)=[\frac 1 x E({\rm Li}_x{\rm NbS}_2)-E({\rm Li-bcc})]/F$,
where $E({\rm Li-bcc})$ is the energy of a Li atom at the bcc Li electrode.
The maximum capacity of bulk NbS$_2$ and its corresponding lithiated composition ${\rm Li}_x{\rm NbS}_2$ can be identified by finding the Li content $x_c$ for which
the voltage drops to zero, $V(x_c)=0$.
Since 
\be V(x_c) - V(x) = \mu(x_c)-\mu(x) = \frac{E_f(x_c)-E_f(x)}{x_c-x},\ee
where $E_f(x)$ is the formation energy per Li atom of  $E({\rm Li}_x{\rm NbS}_2)$. 
thus, the lithium content at full capacity $x_c$ can also be found by minimising the formation energy
\be E_f(x)=E({\rm Li}_x{\rm NbS}_2)-xE({\rm Li\mbox{-}bcc})-E({{\rm NbS}_2}), \ee
where $E({\rm Li}_x{\rm NbS}_2)$ is the total energy per formula unit (f.u.) NbS$_2$  of the lithiated supercell and $E({{\rm NbS}_2})$ is the total energy per f.u. of pristine $2H$-NbS$_2$. All the energies were calculated using DFT, as detailed in the methods section.

The structure of pristine NbS$_2$ has AA$^\prime$ stacking (for stacking nomenclature, please refer to Ref.\cite{tao2014stacking}, with the Nb atoms of one layer stacked on top of the S atoms of the neighbouring layers. The hexagonal interstitial sites of all layers are aligned and therefore, in contrast to graphite, there is no change of stacking for stage I intercalation (for Li$_x$NbS$_2$ with $x$ = 1/4, 1/3, 1/2, 2/3, 1). Some of the most important phases of Li intercalation are shown in Fig. 1.
For LiNb$_2$S$_4$, all the hexagonal sites are occupied. This is different from the intercalated graphite LiC$_6$ where there are no nearest-neighbor occupied hexagonal sites, and possibly is due to the increased screening of NbS$_2$, with a larger density of states at the Fermi level than graphite.

Notably, there is a stable phase for $x$=2 corresponding to a double Li layer intercalation that is not present in graphite. For $2 > x > 1$, the Li atoms start arranging themselves in double layers resembling the stacking of Li-hcp, and the NbS$_2$ layers prefer AB$^\prime$ stacking so as to accommodate the relative shift of the Li layers. The maximum capacity of NbS$_2$, corresponding to the Li$_2$NbS$_2$ phase, is 340.8 mAh/g (Fig. 1). This is smaller than graphene due to the larger masses of Nb and S. However, NbS$_2$ can accommodate 12 times more Li atoms per cell, having a volumetric capacity of 1,364.6 mAh/cm$^3$ (compared to 720.0 mAh/cm$^3$ for graphene). 
The Li$_2$NbS$_2$ phase is expanded by 15\% and 6\% along the $\hat{a}$ and $\hat{c}$ directions, respectively.

\begin{figure*}[h!]
\centering
\includegraphics[width=\textwidth]{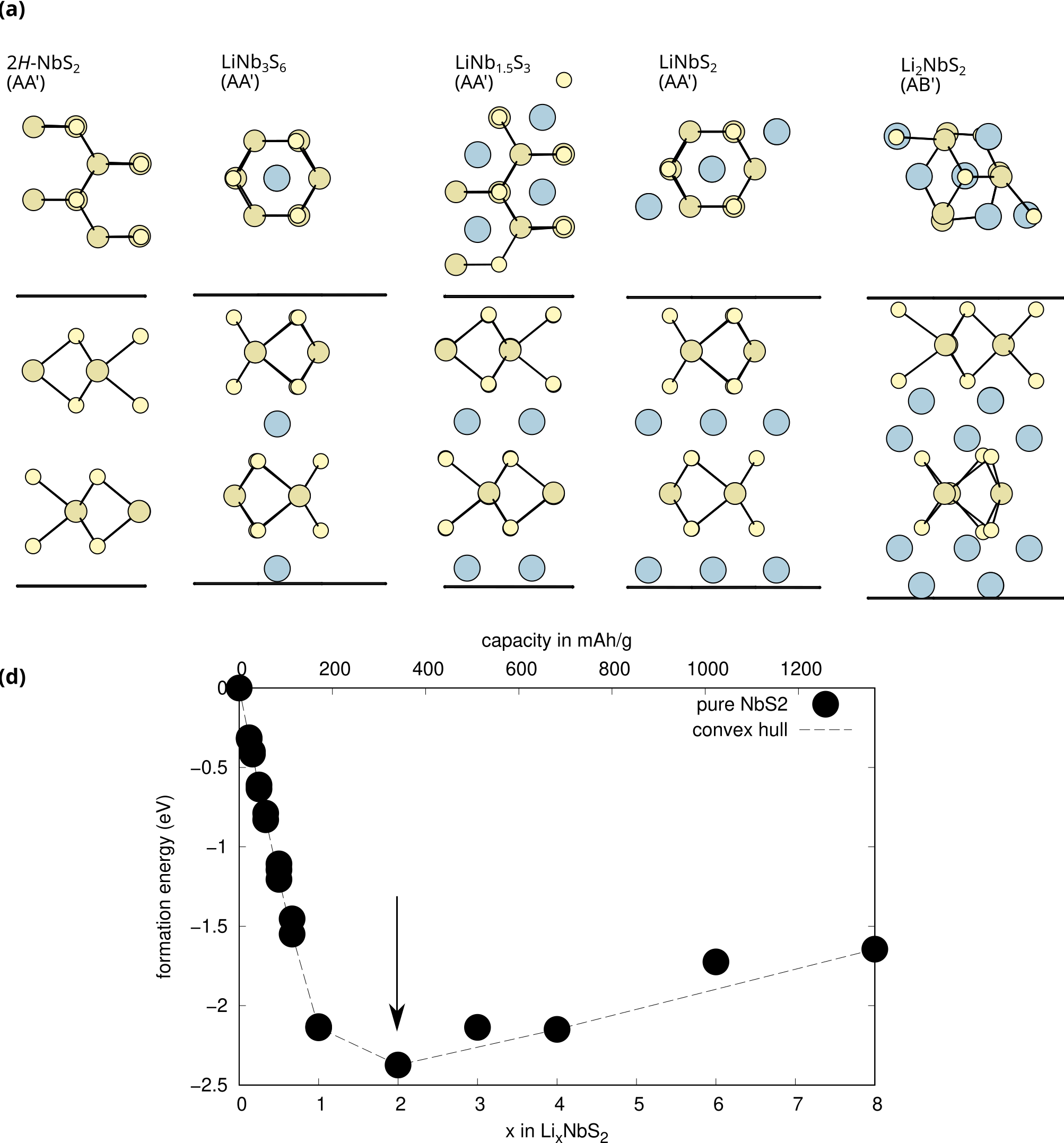}
\caption{Structures of (a) Li-intercalated pristine NbS$_2$ (Li$_x$NbS$_2$), and (b) formation energy vs. Li content for pristine NbS$_2$ (Li$_x$NbS$_2$). Nb, S and Li atoms are represented in dark yellow, light yellow, and blue respectively. The arrow indicates the lowest formation energy, corresponding to maximum capacity. }
 \label{fig:struc}
\end{figure*}

\begin{figure*}[h!]
\centering
\includegraphics[width=\textwidth]{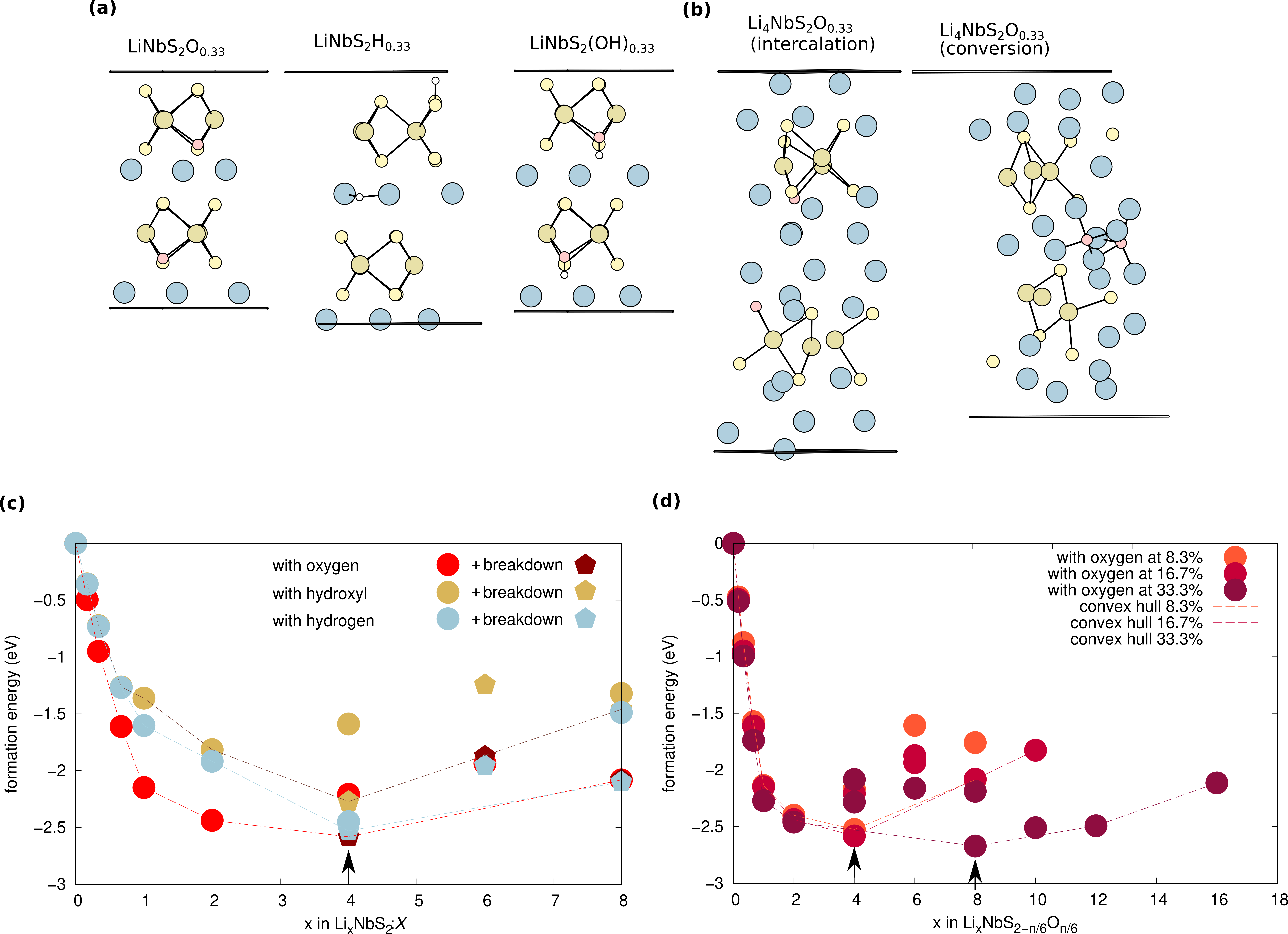}
\caption{  (a) Li-intercalated defective NbS$_2$ with O, OH replacing for S and/or hydrogen doping, for low Li content, and (b) intercalation vs. conversion (alloying) of Li$_x$NbS$_{2-y}$O$_y$, with 2 defects per unit cell and (c) formation energy of partially oxidized lithiated NbS$_2$ (Li$_x$NbS$_{2-y}$O$_y$), as a function of the Li content. Percentages of defects are the percentage of S substitution. Nb, S and Li atoms are represented in dark yellow, light yellow, and blue respectively, O and H atoms in pink and white, respectively. The arrow indicates the lowest formation energy, corresponding to maximum capacity.  }
 \label{fig:struc2}
\end{figure*}

Since the material used in the experiments contained oxygen and hydrogen in various fractions, we have considered the effect that such impurities may have on the capacity. Most importantly, we have found that in the presence of defects, a conversion stage follows the intercalation stage, allowing for higher Li content to be alloyed into NbS$_2$, but with a loss of the layered structure. 

Sulphides  can have oxygen vacancies due to the volatility of sulphur, and these can be filled by oxygen.\cite{hu2019role}
 Oxygen has the same valency as sulphur and therefore replaces sulphur with very little lattice distortion (Fig. 2-a). However, oxygen is more electronegative than sulphur, forming stronger bonds with Li. Therefore,
 there is a competing structure where each of the oxygen atoms exchanges place with a Li, and the NbS$_2$ layers are significantly distorted (Fig. 2-b). This structure is lower in energy than the structure where the oxygen atoms are substitutional (O$_S$). 
 For 16.7\% oxygen concentration (1/6th of the sulphur replaced by oxygen ), the energy gain is 0.37 eV/f.u. for $\rm Li_4NbS_2$, bringing $x=4$ to a lower formation energy than $x=2$ (Fig. 2-c), thus doubling the capacity.

The oxygen presence thus leads to higher lithium intake, leading to partial NbS$_2$ lattice breakdown, similar to what happens in Si anodes. Such conversion-type reaction has also been theoretically predicted for other transition metal chalcogenides.\cite{zhao2019electrochemical} The essential difference in the case of NbS$_2$ is that the presence of defects is necessary to trigger the conversion-type Li insertion for $x$>2.  The formation energy of the lithiated phases, including those that are the product of conversion-type reactions, is shown in Fig.2-b) for an oxygen content of 16\%. Up to $x$ = 2, the succession of phases and their formation energies are nearly unaltered. However, for $x$ = 4, the intercalated phase is 0.4 eV/f.u. higher in energy than the alloyed phase obtained after the reaction of the oxygen atoms with Li, both shown in Fig. 2-b. In the alloyed structure, Li atoms have taken the place of the oxygen atoms. The oxygen atoms are at the inner Li layers, connected to 8 Li atoms each, in a local environment that resembles Li$_2$O.  The maximum intercalation capacity is doubled to 681.6 mAh/g (Fig. 2-c). 

As the oxygen concentration increases up to 33.3\%, the minimum formation energy shifts to $x=8$ in Li$_8$NbS$_{1.33}$O$_{0.67}$ (Fig. 2-d)). The capacity for Li intake for this concentration is 1,459.8 mAh/g. The increase in capacity is also at the expense of conversion of the structure. We found that if the oxygen fraction exceeds the sulphur fraction, the hull curve becomes too deep in energy, and the conversion process can be considered irreversible at normal operating conditions. 

We found that the effect of hydrogen and hydroxyl defects is similar, both triggering covalent bond breaking in the presence of large fractions of Li, and allowing the material to undergo a conversion reaction when present.  A single hydrogen atom prefers to bond to the S atom, forming a sulfhydryl group. The interstitial hydrogen at the Li plane is 1.22 eV higher in energy. However, if a pair of hydrogens are present, with the first in a sulfhydryl group, the second loses its electron to the vicinity of the first sulfhydryl group, becoming an interstitial (Fig. 2-a). For a hydrogen:sulphur ratio of 1:6, we find a higher capacity than for the pristine material (Fig. 2-c). Similar to the case of oxygen, the maximum intercalation is at x = 4 and is accompanied by lattice disruption, with Li taking the place previously occupied by H. Hydroxyl radicals are another possible occurrence at the sulphur site. Similar to oxygen, there is little lattice disruption for x $\leq$ 2. For x = 4, there is a breakdown of the Li lattice, with Li atoms and hydroxyl units exchanging place. This also leads to higher capacity (Fig. 2-c). 

The voltage profile can be derived from the hull energies. The voltage drop between two lithiated phases with Li contents $x_0$ and $x_1$ is given by 
\be V_{x_1}-V_{x_0} =  G({\rm NbLi}_{x_1}{\rm S}_2 - G({\rm NbLi}_{x_0}S_2) /(x_1 - x_0 )       \ee
where the free energy $G$ can be approximated by the formation energy. Setting the intercalated phase at the maximum capacity limit $x = x_{\rm max}$ as the zero voltage, we obtain the average potential as a function of the Li intercalation content shown in Fig. 3. 

\begin{figure*}[h!]
\centering
\includegraphics[width=\textwidth]{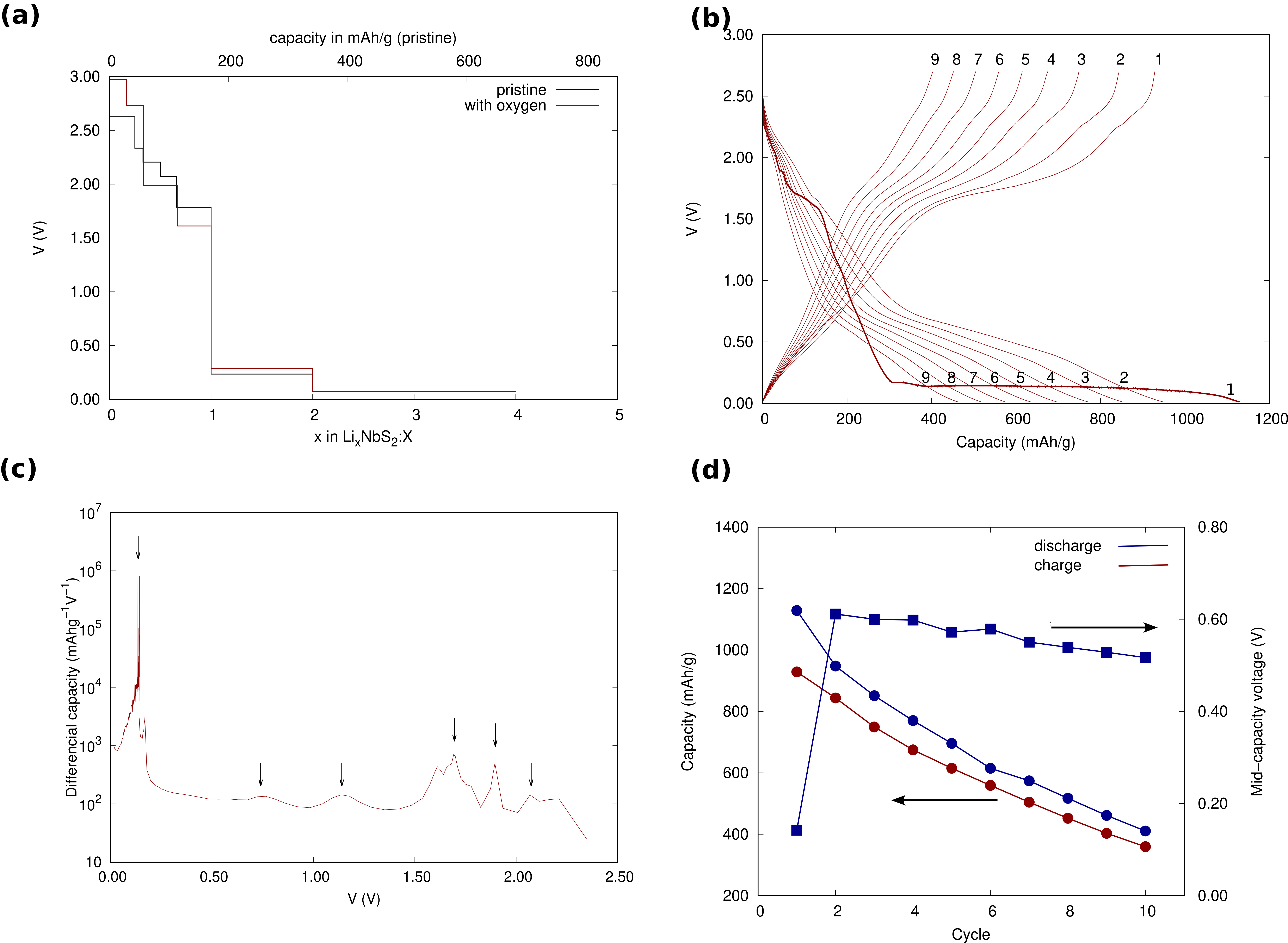}
\caption{(a) Theoretical potential difference with respect to metallic Li; (b) experimental voltage-capacity curves for the first nine charge and discharge cycles (c) derivative of the first-cycle experimental capacity with respect to voltage and (d) degradation of the experimental charge and discharge capacities and mid-capacity voltage. }
 \label{fig:V}
\end{figure*}

To compare with the theoretical calculations, half cells were constructed using Li-metal as the counter and reference electrodes against NbS$_2$-based (99.99\% Niobium Disulfide, American Elements, CAS 12136-97-9) working electrodes. Celgard 2325 and Whatman GF/A were used as the separators and 1.0 M lithium hexafluorophosphate (LiPF$_6$) in ethylene carbonate and dimethyl carbonate (1.0 M LiPF$_6$ in EC/DMC 50/50 (v/v)) as the electrolyte. The working electrodes, with active material mass loading of ~1.3 mg, were made using the ratio composition of 90/5/5 (NbS$_2$/Super-P carbon black/CMC+SBR) tape-casted on Cu foil. Galvanic charge and discharge were performed to obtain the specific average capacity of the NbS$_2$ and cycling performance. The studies were performed with current of 0.02~mA (C-rate of 0.05C) between 0.012 and 2.7~V. 

As presented in Figure 3b, NbS$_2/$Li delivers an initial specific discharge capacity of around 1,130 mAh/g and a first cycling-specific capacity of about 930 mAh/g, with a Coulombic efficiency of 82.3\%. From the second cycle onwards, a systematic specific capacity decay of around 10\% was observed from each cycle to the previous. A stable Coulombic efficiency between 87 and 89\% was found after the first cycle, where was detected. We found similar electrochemical behaviors in our previous work with NbS$_2$-based anode materials\cite{nair-patent}. 

The experimental voltage vs. capacity profile for the first cycle shows several stages with transitions at approximately 2.07, 1.89, 1.69, 1.14, 0.74 and 0.14 V. Even though it is difficult to identify specific stages, the two especially prominent stages at 1.69 and 0.14~V seem to be consistent with the theoretical phase transitions to $x=1$ and to $x=2$, which corresponds to the intercalation of a full Li monolayer or a full Li bilayer.  

A notorious change in the discharge profile was observed between the first and second Galvanic discharges. As the cycles are repeated, the stages are less pronounced, the capacity decreases and the mid-capacity voltage decreases (after the second cycle). This can be due to irreversible reactions of Li with defects, and irreversible increase of the NbS$_2$ interlayer distance. In the presence of impurities and defects, further reaction of Li with the NbS$_2$ layers may also take place, as indicated by our calculations in material containing O and/or H. 

The initial measured capacity of 1,130~mAh/g exceeds the theoretical predictions. The reason for this is that the experiments were carried out with imperfect material, and not single crystals. It is known that large scale defects can lead to an increase in anode capacity, due to Li storage at voids and Li trapping at defects\cite{wang2022introducing,lou2018pseudocapacitive}, and this is a possible reason for the higher experimental capacity.
A variety of disordered carbons have been reported to have significantly higher capacities than graphite\cite{lim2017lithium}, and one of the reasons is the adsorption of Li on both sides of the monolayers and on the edges\cite{gnanaraj2001comparison}. We have performed calculations in monolayer NbS$_2$, showing that the capacity is the double of the bulk capacity, due to Li bilayer adsorption on both sides ($\rm Li_4NbS_2$). Thus, layer disorder and decoupling may also contribute to an increased capacity, similar to disordered graphitic anodes.

After 9 cycles, the capacity is approaching the theoretical estimate for the pristine material, which could possibly be due to the reversible charge and discharge of intercalated Li but not Li reacted with the NbS$_2$ layers and the impurities. 

In summary, we have shown that NbS$_2$ is an anode material with a theoretical capacity up to 340.8 mAh/g in the pristine state. Further, NbS$_2$-based anodes are very robust in the presence of oxygen and hydrogen, which even lead to at least a two-fold increase in capacity. The increased capacity is due to the reaction of Li with the defective NbS$_2$ layers and is possibly associated with irreversibility. 
Experimentally, the NbS$_2$-based anodes were found to have a capacity of 1,130 mAh/g. Two prominent stages at voltages of 1.69 and 0.14~V correspond to the formation of the LiNbS$_2$ and Li$_2$NbS$_2$ phases. 
At maximum capacity, NbS$_2$ has a volume expansion of 42\% upon lithiation, which is much lower than the volume expansion of silicon ($\simeq$ 300\%) \cite{galvez2018simulations}.
We suggest NbS$_2$ as a possible alternative to silicon additives to increase anode capacity.

\paragraph{Acknowledgements}
This research project is supported by the Ministry of Education, 
Singapore, under its Research Centre of Excellence award to the Institute for Functional Intelligent Materials, National University of Singapore
(I-FIM, project No. EDUNC-33-18-279-V12). 
This work used computational resources of the Centre of Advanced 2D Materials (CA2DM), funded by the 
National Research Foundation, Prime Ministers Office, Singapore; and the Singapore National Supercomputing Centre (NSCC).

\section{Methods}
\subsection*{First-principles calculations} 
First-principles calculations were based on the framework of DFT, as implemented in the {\sc Quantum ESPRESSO} package\cite{giannozzi2017advanced}. 
The PBE\cite{PBE} exchange and correlation energy functional was used. Ultra-soft pseudo-potentials\cite{rkkjus} were used for all elements except niobium, for which a norm-conserving Troullier-Martins pseudo-potential was selected\cite{tm-pseudo}. We employed a plane wave basis set with kinetic energy cutoffs of 80~Ry to describe the electronic wave functions. 
The Brillouin zone was sampled using a $\Gamma$-centered 4$\times$4$\times$2 Monkhorst-Pack (MP) grid\cite{mpgrid} for all NbS$_2$ supercell calculations.

\bibliography{refs.bib}
\end{document}